\documentclass[prl,superscriptaddress,showpacs,twocolumn,longbibliography]{revtex4-1}

\usepackage{hyperref}

\usepackage{color}
\usepackage[usenames,dvipsnames]{xcolor}
\usepackage{amsmath,amsthm,amssymb}
\usepackage{graphicx}
\usepackage{epsfig}
\usepackage{dcolumn}
\usepackage{bm}
\usepackage{mathrsfs}
\usepackage{multirow}
\usepackage[all]{xy}
\usepackage{pbox}
\usepackage{verbatim}

\newcommand{\er}[1]{Eq.~\eqref{#1}}
\newcommand{\ers}[2]{Eqs.~(\ref{#1}-\ref{#2})}
\newcommand{\era}[2]{Eqs.~(\ref{#1}) and (\ref{#2})}

\newcommand{\Er}[1]{Equation~\eqref{#1}}
\newcommand{\Ers}[2]{Equations~(\ref{#1}-\ref{#2})}
\newcommand{\Era}[2]{Equations~(\ref{#1}) and (\ref{#2})}

\DeclareMathOperator{\Tr}{Tr}
\def\(({\left(}
\def\)){\right)}
\def\[[{\left[}
\def\]]{\right]}
\def\dd{\text{d}}

\newcommand{\be}{\begin{equation}}
\newcommand{\ee}{\end{equation}}
\newcommand{\ben}{\begin{eqnarray}}
\newcommand{\een}{\end{eqnarray}}
\newcommand{\beq}{\begin{equation}}
\newcommand{\eeq}{\end{equation}}

\newcommand{\EE}{\mathbb{E}}

\newcommand{\Drel}{{\cal D}}  

\begin{document}

\title{Unravelling the large deviation statistics of Markovian open quantum systems}

\author{Federico Carollo}
\affiliation{School of Physics and Astronomy}
\affiliation{Centre for the Mathematics and Theoretical Physics of Quantum Non-Equilibrium Systems,
University of Nottingham, Nottingham, NG7 2RD, UK}
\author{Robert L. Jack}
\affiliation{Department of Applied Mathematics and Theoretical Physics, University of Cambridge, Wilberforce Road, Cambridge CB3 0WA, United Kingdom}
\affiliation{Department of Chemistry, University of Cambridge, Lensfield Road, Cambridge CB2 1EW, United Kingdom}
\author{Juan P. Garrahan}
\affiliation{School of Physics and Astronomy}
\affiliation{Centre for the Mathematics and Theoretical Physics of Quantum Non-Equilibrium Systems,
University of Nottingham, Nottingham, NG7 2RD, UK}

\date{\today}

\begin{abstract}
We analyse dynamical large deviations of quantum trajectories in Markovian open quantum systems in their full generality. We derive a {\em quantum level-2.5 large deviation principle} for these systems, which describes the joint fluctuations of time-averaged quantum jump rates and of the time-averaged quantum state for long times. Like its  level-2.5 counterpart for classical continuous-time Markov chains (which it contains as a special case) this description is both {\em explicit and complete}, as the statistics of arbitrary time-extensive dynamical observables can be obtained by contraction from the explicit level-2.5 rate functional we derive. Our approach uses an unravelled representation of the quantum dynamics which allows these statistics to be obtained by analysing a classical stochastic process in the space of pure states. For quantum reset processes we show that the unravelled dynamics is semi-Markov, and derive bounds on the asymptotic variance of the number of quantum jumps which generalise classical thermodynamic uncertainty relations. We finish by discussing how our level-2.5 approach can be used to study large deviations of non-linear functions of the state such as measures of entanglement.
\end{abstract}

\maketitle 

\noindent {\bf \em Introduction -- }
Practical quantum systems are always coupled to their environments, which means that their dynamics are stochastic.  This is manifested for example by wavefunction collapse and by decoherence.
In such \emph{open quantum systems} one aims to trace out the environment and follow the dynamics of the system 
state~\cite{Plenio1998,Breuer2002,Gardiner2004,Wiseman2009}. 
In many situations, this leads to a density matrix $\rho$ that evolves deterministically in continuous time,
according to a Markovian quantum master equation (QME).  
This dynamics can be understood via a mapping to stochastic quantum trajectories
\cite{Belavkin1990,Dalibard1992,Gardiner1992,Carmichael1993} -- this is called \emph{unravelling} the QME.
An individual quantum trajectory specifies the behaviour of the system conditioned on a time-record of observations (or events)
in the environment.  If the events are {\em quantum jumps} (as for example in the case of photon counting) the trajectories are those of a continuous-time quantum Markov chain, see \cite{Plenio1998,Breuer2002,Gardiner2004,Wiseman2009}.  Averaging over these recovers the QME, but information about their fluctuations requires knowledge about the quantum trajectories.

The state-of-the art  approach for characterising fluctuations in stochastic trajectories uses {\em large deviation principles} (LDPs)~\cite{Giardina2006,Garrahan2007,Lecomte2007,Garrahan2009,Touchette2009,Hedges2009,Jack2010,Chetrite2013,Chetrite2015}. 
This method focuses on rare events in which time-averaged quantities deviate significantly from their typical (ergodic) values. In open quantum systems, LDPs have been used to analyse the counting statistics of quantum jumps~\cite{Garrahan2010,Ates2012,Znidaric2014,Znidaric2014b,Buca2014,Carollo2017,Carollo2018} and of homodyne currents \cite{Hickey2012}. For classical systems, two important recent advances have been the analysis of LDPs for the full statistics of all fluxes  and state occupancies (LDPs at \emph{level 2.5} \cite{Maes2008,Bertini2015b,Barato2015,Hoppenau2016,Bertini2018}), and variational analyses based on \emph{optimal control theory} \cite{Chetrite2015b,Jack2015b}.  Here, we extend these ideas to stochastic quantum trajectories.

In particular, we establish a level-2.5 LDP for quantum jump trajectories, including variational representations of rate functions, based on optimal-control theory.  This framework recovers previous results for the statistics of arbitrary dynamical observables (by using a {\em contraction} principle \cite{Touchette2009}). In addition, it enables several new applications, two of which we consider in detail. First, the level-2.5 LDP allows to derive bounds on the precision of estimation of the (empirical) rates of quantum jumps in quantum reset processes, thus generalising classical \emph{thermodynamic uncertainty relations}~\cite{Barato2015b,Gingrich2016,Pietzonka2017,Garrahan2017,Barato2018}.  Second, the level-2.5 LDP can be used to analyse new kinds of dynamical fluctuations, which are related to non-linear functions of the state; as an example, we consider fluctuations of the bipartite entanglement entropy.

\smallskip

\noindent {\bf \em Average and unravelled dynamics -- }
We consider Markovian open quantum systems in continuous time, where the system density matrix $\rho_t$ evolves according to a QME, 
$\partial\rho_t/\partial t=\mathcal{L}(\rho_t)$. The Lindbladian $\cal L$ \cite{Lindblad1976,Gorini1976} acts on density matrices as \cite{Plenio1998,Breuer2002,Gardiner2004}
\begin{equation}
\mathcal{L}(\cdot) =-i[H,(\cdot)]+\sum_i\left(J_i(\cdot) J_i^\dagger-\frac{1}{2} \{ J^\dagger_i J_i,  (\cdot) \} \right),
\label{L}
\end{equation}
where $J_i$ is a jump operator, and $i=1,2,\dots,m$ identifies the type of quantum jump. For example, different types of jumps might correspond to emitted photons with different frequencies.  We write $[A,B]=AB-BA$ for the commutator of two operators and $\{A,B\}=AB+BA$ for their anti-commutator.

Our approach is based on {\em unravelling} the dynamical evolution described by the QME in terms of quantum jump trajectories \cite{Plenio1998,Breuer2002,Gardiner2004}. Each trajectory is the stochastic evolution of a pure state, which we denote at time $t$ by the density matrix $\psi_t$, with $(\psi_t)^2=\psi_t=\psi_t^\dag$ and $\Tr \psi_t = 1$. 
%
The matrix $\psi_t$ evolves according to a Belavkin {\em stochastic differential equation} (SDE) \cite{Belavkin1990}, 
\begin{equation}
d{\psi}_t=\mathcal{B}({\psi}_t)dt+\sum_i\left(\frac{\mathcal{J}_i({\psi}_t)}{\Tr[\mathcal{J}_i({\psi}_t)]}-{\psi}_t\right)dn_{it}\, ,
\label{BL}
\end{equation}
where 
$$
\mathcal{B}({\psi})
=-iH_{\rm eff}\psi+i\psi H_{\rm eff}^\dagger -\psi\Tr(-iH_{\rm eff}{\psi}+i{\psi} H_{\rm eff}^\dagger) \; ,
$$ 
with $H_{\rm eff}=H-\frac{i}{2}\sum_iJ^\dagger J_i$
being the (non-Hermitian) effective Hamiltonian, and $\mathcal{J}_i(\psi) = J_i{\psi} J^\dagger_i$.  
Formally, the ``noise increment'' $dn_{it}$ is equal to one if a jump of type $i$ takes place between times $t$ and $t+dt$ or zero otherwise. The average of $dn_{it}$ is $\Tr[\mathcal{J}_i({\psi}_t)]dt$, and the noise increments obey ``Ito rules'' $dn_{it} dn_{jt} = \delta_{ij} dn_{it}$ \cite{Breuer2002,Gardiner2004}. Two standard results are (i) the Belavkin SDE maintains $\psi_t$ as a pure state, and (ii) for appropriate initial conditions on $\psi_t$, the density matrix can be recovered by averaging $\psi$ over the noise realisations: $\rho_t = \mathbb{E}[\psi_t]$.  Hence, all quantum observables can be computed as classical expectation values for the unravelled process.

\smallskip

\noindent {\bf \em Quantum-classical correspondence and master equation for unravelled dynamics -- } 
\Er{BL} represents the quantum Markov chain via a classical SDE in the space of pure states $\psi$. Let $P_t(\psi)$ be the probability density for $\psi$ at time $t$, in analogy with classical stochastic processes.  Then
\begin{multline}
\partial_t P_t(\psi)
=-\operatorname{div} \left[\mathcal{B}(\psi)P_t(\psi)\right] \\
+\sum_i\int d\psi'\, \left[P_t(\psi')w_i(\psi',\psi)-P_t(\psi)w_i(\psi,\psi')\right],
\label{FP}
\end{multline}
where 
\be
w_i(\psi,\psi')=\Tr\left[\mathcal{J}_i(\psi)\right]\delta\left(\psi'-\frac{\mathcal{J}_i(\psi)}{\Tr\left[\mathcal{J}_i(\psi)\right]}\right)
\label{wi}
\ee
is the rate for transitions from $\psi$ to $\psi'$ due to quantum jump $i$. Precise definitions of the quantities in \er{FP} are given in \cite{SM}. We call \er{FP} the unravelled dynamics quantum master equation (UQME). 

Physically, \ers{BL}{FP} have a simple meaning: the pure state $\psi_t$ evolves deterministically along paths specified by the operator $\cal B$, but this deterministic evolution is punctuated at random times by jumps, specified by ${\cal J}_i$. The probability $P_t$ evolves according to the UQME, and at long times it tends to the stationary solution $P_\infty(\psi)$.  We assume that this solution is unique, which is the case in most physical applications, see also~\cite{Benoist2018}.

We summarise this quantum-classical mapping (or unravelling) as: (i) $\psi_t$ is the (stochastic) {\em  position} in Hilbert space which evolves according to the SDE~\eqref{BL}; (ii) the state $\rho_t$ corresponds to the {\em average position} and evolves according to the QME~\eqref{L} 
\footnote{Note that the Lindblad equation \eqref{L} is a closed equation for the evolution of $\rho=\mathbb{E}[\psi]$.  In  classical stochastic processes, expectation values do not generally have closed evolution equations.  The Belavkin SDE \er{BL} is special in that it leads to a closed evolution for the average of $\psi_t$.}; (iii) the master equation for the stochastic process $\psi_t$ is the UQME.

\smallskip

\noindent {\bf \em Level 2.5 LDP for unravelled dynamics --} 
We derive a LDP at level 2.5 for the unravelled process by generalising the classical result to systems that evolve by a combination of continuous deterministic evolution and discrete (random) jumps, cf.\ \ers{BL}{wi}. The large deviation (LD) theory of stochastic dynamics is concerned with the behaviour of observables that are time-integrated over trajectories, for some long time $\tau$ \cite{[For a simple review see ]Garrahan2018}. At level 2.5 these observables fall into two main classes \cite{Maes2008,Bertini2015b,Barato2015,Hoppenau2016,Bertini2018}: {\em empirical fluxes} ${q}^i_\tau(\psi,\psi')$, corresponding to the number of jumps from $\psi$ to $\psi'$ per unit time in a trajectory (i.e., empirical transition rates), and the \emph{empirical measure} $\mu_\tau(\psi)$, corresponding to the fraction of time that the system spends in $\psi$. Their (steady-state) averages over trajectories are $\EE[\mu_\tau(\psi)] = P_\infty(\psi)$ and $\EE[{q}^i_\tau(\psi,\psi')] = P_\infty(\psi)w_i(\psi,\psi')$.  
The level-2.5 LDP quantifies the (small) probability that $\mu_\tau$ and $q_\tau$ differ from their average values: as $\tau\to\infty$ then
\be
\mathrm{Prob}[(\mu_\tau, q_\tau) \approx (\mu,q)] \simeq \exp\left( -\tau I_{2.5}[\mu,q] \right)
\label{ldp}
\ee
where $I_{2.5}[\mu,q] $ is the level-2.5 rate functional.  

To obtain a formula for $I_{2.5}$, we define a \emph{controlled stochastic process}, in which the transition rates $w_i$ are replaced by auxiliary rates \cite{Chetrite2015b,Jack2015b}
\be
{w}^{\rm A}_i(\psi,\psi') = A_i(\psi) w_i(\psi,\psi') 
\label{wA}
\ee
where $A_i$ is a (positive) rescaling factor.
The steady state probability density for this controlled process is denoted by $P_\infty^{\rm A}(\psi)$, which may (in principle) be obtained as the steady-state solution of a suitable UQME.  By considering cumulant generating functions for $\mu,q$ and performing a Legendre transformation (see \cite{SM} for details) we obtain
\be
I_{2.5}[\mu,q] = \inf_{A} \, {\cal I}[w^A|w]
\label{L2.5-inf}
\ee
where the infimum is taken over all possible choices of the rescaling factors $A$ such that the controlled process realises the rare values of $\mu,q$: that is $P_\infty^{\rm A}(\psi)=\mu(\psi)$ and $P_\infty^{\rm A}(\psi)w^A_i(\psi,\psi')=q^i(\psi,\psi')$.  If there is no choice for $A$ that satisfies this constraint then $I_{2.5}[\mu,q]=\infty$.  The controlled process that corresponds to the minimiser in (\ref{L2.5-inf}) is the \emph{optimally-controlled process}.  The quantity to be minimised is a relative entropy
\be
{\cal I}[w^A|w] = \int d\psi d\psi'  P^{\rm A}_\infty(\psi) \sum_i  \Drel \!\left[w_i^A(\psi,\psi')\big|w_i(\psi,\psi')\right]
\label{L2.5-ent}
\ee
where 
\be
\Drel(x|y) = x \log(x/y) - x + y
\ee
Using the formulae for $w_i^A$ and the UQME \eqref{FP}, one has
\be
I_{2.5}[\mu,q] = \int d\psi d\psi'\sum_i  \Drel \!\left[q^i(\psi,\psi')\big |\mu(\psi) w_i(\psi,\psi')\right]
\label{L25-direct}
\ee
which is valid as long as the continuity constraint 
\be
\operatorname{div}[\mathcal{B}(\psi)\mu(\psi)] = \sum_i\int d\psi' \big[ q^i(\psi',\psi) 
- q^i(\psi,\psi')\big]
\label{continuity}
\ee
is satisfied for all $\psi$.  Otherwise $I_{2.5}=\infty$.  \Era{L25-direct}{continuity} are analogous to the classical theory of LDs at level 2.5, but generalised to quantum Markovian dynamics, in which the system evolves deterministically between its (random) jumps 
\footnote{The quantum level 2.5 LDP, \era{L25-direct}{continuity}, contains the one for {\em classical} Markov processes as a special case, obtained when $H=0$ with all the $J_i$ being rank-1 operators mapping configuration states. In this case $H_{\rm eff}$ is diagonal, pure states $\psi$ corresponds to individual configurations in the classical basis, and $\mathcal{B}({\psi})$ vanishes. Correspondingly, \era{L25-direct}{continuity} reduce to the usual classical level 2.5 description.}.
\Ers{ldp}{continuity} show how large deviations for open quantum dynamics can be analysed at level 2.5, and they establish a {\em variational principle} for the rate function. This is the first main result of the paper. We now discuss its consequences.

\smallskip

\noindent {\bf \em Contraction to level 1 LDPs and quantum Doob transform --}
As in the classical case, the level 2.5 LDP rate function is given by an {\em explicit} expression, in terms of empirical fluxes and the empirical measure, cf.\ \ers{ldp}{continuity}.  This LDP is {\em complete} in the sense that the rate function for any linear combination of the empirical fluxes and measure can be derived from~\er{L25-direct} by the contraction principle of large deviation (LD) theory.  For example, the number of quantum jumps of type $i$ per unit time is the integral of the empirical flux over all possible initial and final states: ${Q}^i_\tau = \int d\psi d\psi' {q}^i_\tau(\psi,\psi')$. These fluxes obey an LDP (known as ``level 1'' \cite{Touchette2009}), $\mathrm{Prob}[{Q}_\tau \approx Q] \simeq e^{-\tau I_1(Q)}$. The rate function $I_1$ can be obtained by contraction from level 2.5: that is,
$I_1(Q) = \min_{\mu,q: Q} I_{2.5}[\mu,q]$, where the minimisation is over all $(\mu,q)$ such that the jump rate is $Q$. 

A second important result for level-1 statistics that can be recovered from our level 2.5 approach is the {\em quantum Doob} transformation \cite{Garrahan2010,Carollo2018}. This states that there is an auxiliary quantum process for which the rare events in $\mathrm{Prob}[{Q}_\tau \approx Q]$ become typical. The derivation consists of three main steps: first, the variational characterisation of $I_{2.5}$ in \er{L2.5-inf} provides an auxiliary process on the space of pure states, which optimally realises the fluctuation $(\mu,q)$; second, applying a similarity transformation to $\psi_t$ yields a new set of quantum stochastic trajectories; third, one shows that these trjaectories are an unravelled representation of the Doob-transformed dynamics. For details see \cite{SM}.  

These results show that the quantum level-2.5 LDP (\ref{ldp}) can be used to recover existing results that are usually calculated through tilted Lindbladian methods \cite{Garrahan2010,Carollo2018}.  However, the level-2.5 LDP contains much more information about the dynamics than the tilted Lindbladian.  As well as fluctuations in the quantum jump rates, it also describes fluctuations of the empirical measure $\mu_\tau$.  Furthermore it provides the variational principle (\ref{L2.5-inf}). These open the door to a range of new studies. We discuss two such directions below.

\smallskip 

\noindent {\bf \em Application 1: Fluctuation bounds in quantum reset processes --}
Classical level 2.5 LDPs have been used to derive lower bounds on the size of fluctuations of currents and fluxes, relating them to entropy production and dynamical activity - these are called ``thermodynamic uncertainty relations'' (TURs) \cite{Barato2015b,Gingrich2016,Pietzonka2017,Garrahan2017,Barato2018}. We now use the variational formula (\ref{L2.5-inf}) to obtain similar bounds for open quantum dynamics. 

We restrict our analysis to \emph{quantum reset processes}, in which each jump operator projects the system into a specific state:  ${\cal J}_i(\psi)= f(\psi) \varphi_i$ where $f$ is a scalar function, and the pure state $\varphi_i$ is independent of $\psi$. In this case, the steady-state distribution $P_\infty$ is supported on a set of $m$ deterministic paths.  It follows that the statistics of jumps can be described by a classical semi-Markov process -- the time between jumps is in general a non-exponential random variable with a distribution that depends on the end-point of the previous jump (but not on the previous history of the process). For a system that makes a jump of type $i$ at $t=0$, the probability that its next jump is of type $j$ and occurs at time $t$ is $p_{ij}(t) = \Tr\left( {J_j}^\dag J_j e^{-iH_{\rm eff}t}\varphi_ie^{iH^\dagger_{\rm eff}t} \right)$.  The marginal probability that this jump is of type $j$ is $R_{ij}=\int_0^\infty dt\, p_{ij}(t)$, the average time for such a jump is $\tau_{ij}=R_{ij}^{-1} \int_0^\infty dt\, t p_{ij}(t)$ and its variance $\sigma^2_{ij}=R_{ij}^{-1} \int_0^\infty dt\, (t-\tau_{ij})^2 p_{ij}(t)$

The statistics of jumps in a quantum reset process are fully determined by the $p_{ij}(t)$.  Moreover, following (\ref{wA}), an auxiliary process can be constructed with an arbitrary distribution $\hat p_{ij}(t)$.   Given such a process whose mean jump rates are $Q^i$, one has from (\ref{L2.5-inf}) that
\beq
I_1(Q) \leq \sum_{ij} Q^i \int_0^\infty dt \, \hat p_{ij}(t) \log \frac{\hat p_{ij}(t)}{p_{ij}(t)}
\label{I1-p}
\eeq
where the explicit relation between $Q$ and $\hat p_{ij}(t)$ is given in \cite{SM}. \Er{I1-p} provides a lower bound on the probability of the rare values of the $Q^i$ (with the optimal process saturating the bound and giving the exact $I_1$). 

The above result can be used to establish a general bound on the variance of the empirical rates, which generalises the classical TURs, as follows (for details see \cite{SM}). We choose the $\hat p_{ij}$ such that all jump times are rescaled uniformly from those of the typical process, $\hat \tau_{ij} = \tau_{ij} / a$, where $a$ is a constant, while the marginal probabilities remain the same, $\hat R_{ij}=R_{ij}$. This can be achieved with $\hat p_{ij} = v_{ij} p_{ij}(t) e^{-u_{ij}t}$ by an appropriate choice of  $v_{ij}$ and $u_{ij}$. The jump counts are also rescaled uniformly, $Q_i = a \, \bar Q_i$, where $\bar Q_i = \EE[Q_i]$ are those of the process $p_{ij}(t)$. Taking $a = 1 + \delta$ with $\delta\ll1$, \er{I1-p} gives
\be
I_1[(1 + \delta) \bar Q] \leq \frac{1}{2} \chi \delta^2 + O(\delta^3)
\label{I1-Qa}
\ee
with $\chi= \sum_{ij} \bar Q_i R_{ij} \tau_{ij}^2/\sigma_{ij}^2$.
This result provides an uncertainty bound for any linear combination of the empirical jump rates, ${\cal Q}_b = \sum_i b_i Q_i$.  That is,
\be
\frac{{\rm var}({\cal Q}_b)}{\bar {\cal Q}_b^2} \geq \frac{1}{\tau \chi}
\label{var}
\ee
where $\bar {\cal Q}_b = \EE[{\cal Q}_b] = \sum_i b_i \bar Q_i$. \Er{var} is a bound on the precision with which ${\cal Q}_b$ can be estimated, and is thus a TUR for quantum reset processes. In the case where the jump probabilities $p_{ij}(t)$ are exponential - corresponding to a classical jump process - \er{var} reduces to the existing classical TUR for counting observables \cite{Garrahan2017}, as $\tau_{ij} = \sigma_{ij}$ giving $\chi = \sum_{ij} \bar Q_i R_{ij}$ which is the average activity. In the open quantum case one may achieve more precise estimates because the bound on precision depends on the reweighed sum in $\chi$. When $\sigma_{ij} < \tau_{ij}$, that is sub-Poissonian, the more precise jump times can lead to less uncertainty in \er{var}. Similar enhancement in precision can occur for example in classical systems with time-periodic dynamics \cite{Barato2018b} or in the presence of magnetic fields \cite{Macieszczak2018}. In our case it is related to the possibility of antibunching of quantum jumps \cite{Plenio1998}.

To illustrate the quantum TUR we consider a very simple system: a single particle which can occupy two sites, Fig.\ 1(a). The Hamiltonian is $H = \Omega \left( |10\rangle \langle 01| + |01\rangle \langle 10| \right)$ which generates coherent hopping at frequency $\Omega$. There is dissipation in the form of \emph{incoherent} hopping, with jump operators $J_{\rm L} =\sqrt{\gamma} |10\rangle \langle 01|$ and $J_{\rm R} = \sqrt{\gamma} |01\rangle \langle 10|$. This is a quantum reset process of the kind described above, with reset states $\varphi_{\rm L} = |10\rangle \langle 10|$ and $\varphi_{\rm R} = |01\rangle \langle 01|$. For the observable ${\cal Q}_b$ we consider the flux due to jumps into $\varphi_L$, so ${\cal Q}_b = Q_{\rm L}$. The corresponding rate function $I_1(Q_{\rm L})$ can be computed exactly from the tilted Lindbladian \cite{Garrahan2010}, full (black) curve in Fig.\ 1(b). Using again the ansatz $\hat p_{ij}(t) \propto e^{-ut} p_{ij}(t)$, \er{I1-p} yields a bound on the whole rate function, dashed (red) curve in Fig.\ 1(b). The inset to Fig.\ 1(b) shows the TUR bound \er{var} close to the minimum of $I_1$.

\begin{figure}[t]
\centering
\includegraphics[width=\columnwidth]{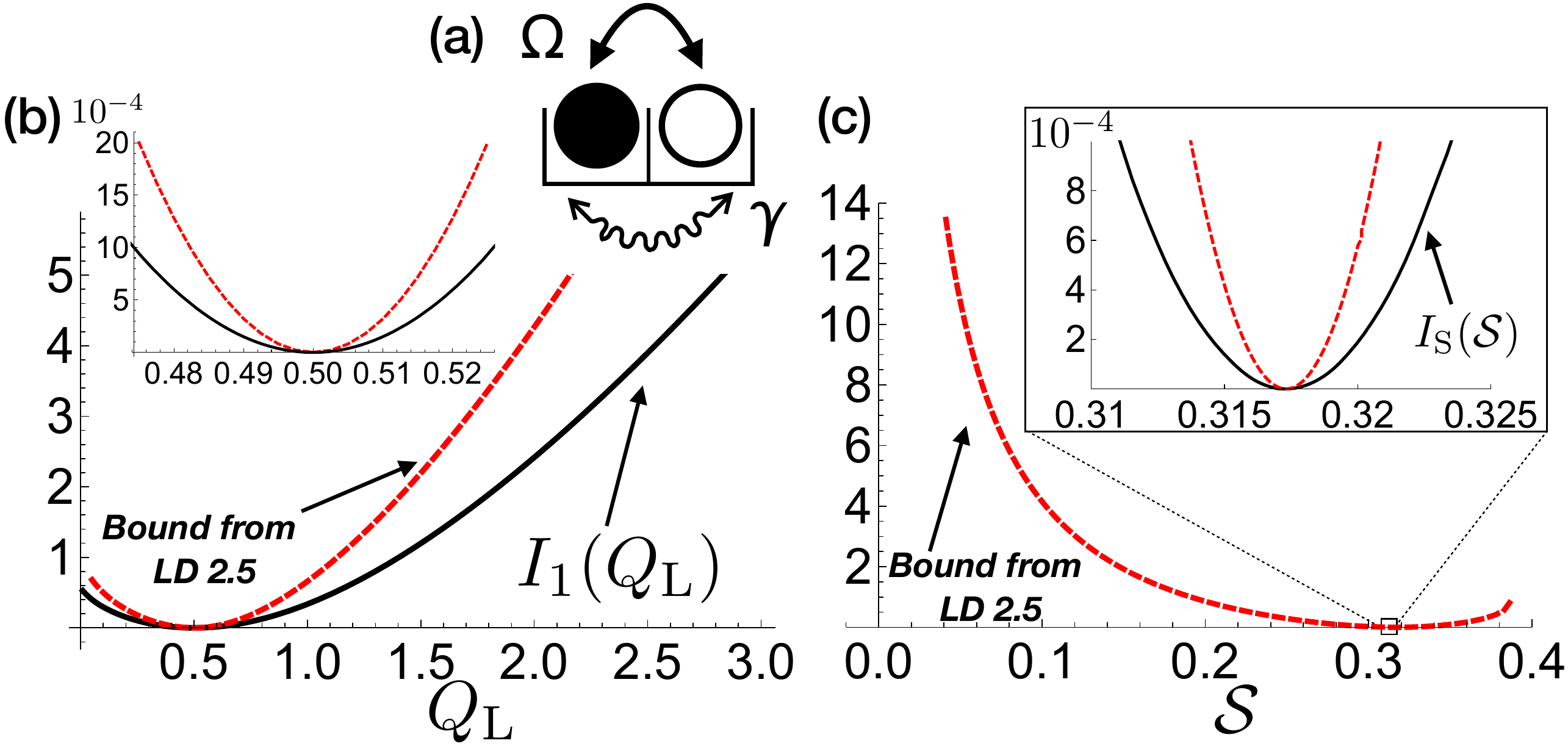}
\caption{(a) Example of a quantum reset process: single particle subject to coherent and dissipative hopping.  (b)~Exact rate function $I_1$ for the empirical flux $Q_{\rm L}$ (full black curve) and  associated bound from \er{I1-p} (dashed red). Inset: behaviour close to the mean corresponding to the TUR \er{I1-Qa}. (c)~An estimate (upper bound) of the rate function for the entanglement, from \er{L2.5-inf}.  The inset shows the behaviour close to the mean, and an associated quadratic bound analogous to (\ref{I1-Qa}).  The largest possible value for ${\cal S}_\tau$ is $\approx 0.3863$, which is achieved by long trajectories that have no incoherent jumps.}
\label{Fig1}
\end{figure}

\smallskip

\noindent {\bf \em Application 2: Statistics of time-integrated entanglement -- }
The level 2.5 LDP (\ref{ldp}) describes the joint fluctuations of $(\mu_\tau, q_\tau)$ - we have concentrated so far on its implications for the statistics of empirical jump rates.  However, the extra information in \er{ldp} may also be exploited to obtain the {\em statistics of nonlinear functions} of the state. A prominent example is the entanglement entropy \cite{Horodecki2009}. 

Consider a bipartite system, where $\psi_t$ is the (pure) state of the whole system at time $t$. The entanglement entropy $S_E$ between parts $A$ and $B$ is $S_E(\psi_t) = -\Tr_A \omega(\psi_t) \log \omega(\psi_t)$, where $\omega(\psi_t)=\Tr_B \psi_t$ is the reduced state in partition $A$, and $\Tr_{A,B}$ denote partial traces over parts $A$ and $B$. In open quantum systems then $\psi_t$ is a random (fluctuating) quantity, as is the nonlinear function $S_E(\psi_t)$.
In particular, the {\em empirical entanglement entropy} (i.e., the time average over a trajectory) ${\cal S}_\tau = \tau^{-1} \int_0^\tau dt\, S_E(\psi_t)$ obeys an LDP for large time: $\mathrm{Prob}[{\cal S}_\tau \approx {\cal S}] \simeq \exp\left[ -\tau I_{\rm S}({\cal S}) \right]$ where $I_{\rm S}$ can again be obtained by contraction from \er{ldp} or by the variational formula \er{L2.5-inf}, restricted to an auxiliary processes where the (mean) entanglement is $\cal S$.  In contrast to the statistics of quantum jumps \cite{Garrahan2010}, the fluctuations of ${\cal S}_\tau$ {\em cannot} be obtained by spectral analysis of a tilted Lindbladian. The application of the quantum level 2.5 LDP \er{ldp} to the statistics of nonlinear functions of the state is the third main result of this paper. 

Fig.\ 1(c) shows the behaviour of the empirical entanglement in the example system of Fig.\ 1(a).  In this example the Lindbladian is {\em unital} \cite{Breuer2002}, so the stationary state density matrix is the identity. As such, the average state has no entanglement for long times. In contrast, the unravelled state $\psi_t$ is typically entangled: the evolution with $H_{\rm eff} = H - i \gamma /2$ between jumps generates entanglement due to coherent hopping, while the dissipative jumps return $\psi_t$ to the product states $\varphi_{L,R}$, which resets the instantaneous entanglement to zero. In Fig.\ 1(c) we show the rate function of $S_\tau$ as estimated numerically by direct simulation of the quantum jump process (full black curve) and the corresponding bound (dashed red) obtained from \er{L2.5-inf} using a similar ansatz as for Fig.\ 1(b). Computation  of the rate function $I_{\cal S}$ would require exact solution of the variational problem \er{L2.5-inf}: here we present a bound that  
applies for all values of ${\cal S}$, including those that are very rare (see \cite{SM} for details).  In general the exact computation of rate functions will be difficult, but the possibility to bound them with simple ansatze make the level 2.5 approach - as in the classical case - both useful and practical.

\smallskip

\noindent{\bf \em Outlook --} The level 2.5 method presented here for quantum Markov chains, can also be formulated for other unravellings, such as those based on homodyne detection described by quantum Wiener processes \cite{Breuer2002,Gardiner2004}. Another interesting extension is to discrete time quantum Markovian dynamics. A possible application of such formulation would be to study the statistics of entanglement, out-of-time-order correlators and operator spreading in random unitary circuits \cite{Nahum2017,Nahum2018,Keyserlingk2018}. In analogy with classical level 2.5 LDPs, the method here can also be extended to time-periodic dynamics, with possible application to periodically driven (Floquet) quantum systems \cite{Moessner2017}. We hope to report on some of these extensions in the near future.

\smallskip

\noindent{\bf \em Acknowledgements --} 
This work was supported by EPSRC grants no.\ EP/M014266/1 (JPG) and EP/N03404X/1 (FC and JPG), and by the European Research Council under the European Union's Seventh Framework Programme (FP/2007-2013) / ERC Grant Agreement No. 335266 (ESCQUMA) (FC). JPG  wishes to thank Madalin Guta for fruitful discussions.  


\bibliographystyle{apsrev4-1}
\bibliography{QL25}

\newcommand{\lind}{1}
\newcommand{\belav}{2}
\newcommand{\uqme}{3}
\newcommand{\wrates}{4}
\newcommand{\wA}{6}
\newcommand{\contin}{11}
\newcommand{\ratefn}{10}
\newcommand{\ratevar}{8}

\newcommand{\citeDoob}{18,24}
\newcommand{\citeHugo}{17}
\newcommand{\citeJuanpeFCS}{18}

\onecolumngrid
\newpage

\renewcommand\thesection{S\arabic{section}}
\renewcommand\theequation{S\arabic{equation}}
\renewcommand\thefigure{S\arabic{figure}}
\setcounter{equation}{0}

\begin{center}
{\Large Unravelling the large deviation statistics of Markovian open quantum systems: Supplemental Material}
\end{center}

\tableofcontents

\section{General theory}

\subsection{Unravelled quantum master equation} 

This section summarises the derivation of (\uqme) and defines the objects that appear in it.  We work in a Hilbert space of dimension $n$ and we use a fixed orthonormal basis, so the pure-state density matrix $\psi$  is of size $n\times n$ with complex elements.  Mathematically, one may write $\psi\in\mathbb{C}^{n\times n}$.  Consider a general function $f$ that depends on $\psi$. Given an initial pure state $\psi_0$, we use (\belav) to calculate the time-derivative of the average of $f(\psi_t)$ at $t=0$.  The result may be written as
\be
\partial_t \mathbb{E}_{\psi_0}[f(\psi_t)] = {\cal W}[f(\psi_0)]
\label{equ:W-dt}
\ee
where, on the right hand side, $\cal W$ is an operator such that ${\cal W}[f]$ is a function of $\psi$, and this function is evaluated at $\psi_0$.  The operator ${\cal W}$ depends on the $\cal B$ and the $w_i$ that appear in (\belav): evaluating the expectation value in (\ref{equ:W-dt}) gives 
\be
\mathcal{W}[f(\psi_0)]= \mathcal{B}[\psi_0] \cdot \operatorname{grad}f(\psi_0)  + \sum_{i}\int d\psi'\, w_i(\psi_0,\psi')\left[f(\psi')-f(\psi_0)\right]\, ,
\label{equ:Wf}
\ee
In this equation, the integral runs over all matrices, in a sense to be defined below.  The meaning of the first term on the right hand side is that  ($\operatorname{grad}f$) is a gradient and the dot indicates an inner product, which are both to be interpreted in the space of matrices, that is $\mathbb{C}^{n\times n}$.  Specifically, for any function $f$ and any $n\times n$ matrix  $X$,
$$
X \cdot \operatorname{grad} f =\sum_{ij} X_{ij} \frac{\partial f}{\partial \psi_{ij}} \, .
$$
Here the matrix elements $\psi_{ij}$ are complex numbers, and $f$ is to be interpreted as a function that depends on all the matrix elements (of which there are $n^2$).

The next step is to consider the derivative $\partial_t \mathbb{E}[f(\psi_t)]$ in cases where the initial matrix $\psi$ is drawn from some initial probability distribution.  Let $P_0$ be the probability density associated with this initial condition, so the probability that $\psi_0$ is within some set $\cal A$ is
$
\int_{\cal A} P_0(\psi) {\rm d}\psi .
$
Here, $\cal A$ is a subset of $\mathbb{C}^{n\times n}$, and the integration measure is ${\rm d}\psi = {\rm d}\psi_{11}\ldots{\rm d}\psi_{nn}$ which means each matrix element is interpreted as a separate complex variable.  The integral in (\ref{equ:Wf}) is interpreted in the same sense.  In cases (such as that one) where the domain of integration is omitted then it is assumed to be the full space, which is $\mathbb{C}^{n\times n}$.

For this theory to make physical sense, we know that the initial distribution $P_0$ must be entirely supported on pure-state density matrices: for example, the probability for any matrix with $\psi\neq\psi^\dag$ is zero.  Hence, $P_0(\psi)\propto \prod_{i\leq j} \delta(\psi_{ij}-\psi_{ji}^*)$ where $^*$ denotes the complex conjugate; there are also other constraints on $\psi$,  for example $\psi^2=\psi$ and $\Tr(\psi)=1$. 

Since we have defined a gradient operator and an integration measure, we can perform integration by parts.  For any matrix-valued function $X=X(\psi)$ with matrix elements $X_{ij}(\psi)$ we define the divergence as
$$
\operatorname{div}[X(\psi)]=\sum_{ij}\frac{\partial X_{ij}(\psi)}{\partial \psi_{ij}}\, ,
$$
This is a complex number (dependent on $\psi)$.
Hence, for any scalar function $f$ and any matrix-valued function $Y$ then we may derive an integration-by-parts formula
$$
\int {\rm d}\psi P(\psi) Y(\psi) \cdot  \operatorname{grad} f(\psi)  = - \int  {\rm d}\psi f(\psi) \operatorname{div}[ P(\psi) Y(\psi) ]
$$
There are no boundary terms since $P$ is finite only if all matrix elements are finite, and each matrix element is integrated over the whole complex plane.
Now multiply (\ref{equ:Wf}) by $P_0(\psi_0)$ and integrate over $\psi_0$; then integrate by parts, and insist that the resulting equation is true for all functions $f$.  This yields 
(\uqme) of the main text, which can be written as $\partial_t P = {\cal W}^*[P]$, where ${\cal W}^*$ is the adjoint of $\cal W$, in the sense that $\int P(\psi) {\cal W}[f(\psi)] \dd\psi = \int {\cal W}^*[P(\psi)] f(\psi) \dd\psi$ for all admissible $f,P$.  Note that  (\belav) preserves $\psi_t$ as a pure state density matrix and (\uqme) is derived from (\belav): it follows that the solution $P_t$ of (\uqme) is entirely supported on pure-state density matrices, as long this property holds for $P_0$.

%
%

\subsection{The level 2.5 functional for quantum Markovian unravelled dynamics}

This section derives the level-2.5 rate function $I_{2.5}[\mu,q]$ that is given in Eq.(\ratefn) of the main text, and its variational representation (\ratevar).  This is achieved by tilting the generator (\ref{equ:Wf}) of the unravelled dynamics.  Define
%
\begin{equation}
\mathcal{W}_{\alpha,\beta}[f(\psi)]=\sum_{i}\int d\psi'\, w_i(\psi,\psi')\left[e^{-\alpha_i(\psi,\psi')}f(\psi')-f(\psi)\right]+\operatorname{grad}f(\psi)\cdot \mathcal{B}[\psi]-\beta(\psi)f(\psi)\, ,
\label{tiltdual}
\end{equation}
with $\alpha_i(\psi,\psi')$ being a field conjugated to the fluxes from the state $\psi$ to the state $\psi'$ through the $i$th jump ($q_{\tau}^i(\psi,\psi')$), while $\beta(\psi)$ is a field conjugated to the empirical measure $\mu_\tau(\psi)$. In the following, we will drop the explicit $\tau$ dependence on the quantitites, always keeping in mind that the derivation below is valid in the large $\tau$ limit. 

Consistent with our assumption that the UQME has a unique steady-state solution, we assume that the (tilted) generator $\mathcal{W}_{\alpha,\beta}$ has a unique (real) largest eigenvalue $\Theta(\alpha,\beta)$, associated with the left/right eigenfunctions $f_{L}(\psi)/f_{R}(\psi)$.  Then the level $2.5$ rate function for the empirical measure and fluxes can be obtained by means of the following Legendre transform:
\begin{equation}
\begin{split}
I_{2.5}[\mu,q]&=\sup_{\alpha,\beta}\left[F(\alpha,\beta)\right]\, ,\\
\mbox{ with }F(\alpha,\beta)=-\sum_i\int d\psi &d\psi'\, q^i(\psi,\psi')\,\alpha_i(\psi,\psi')-\int d\psi\,\beta(\psi)\mu(\psi)-\Theta(\alpha,\beta)\, .
\end{split}
\label{LegTransf}
\end{equation}
Since $f_L(\psi)$ and $f_R(\psi)$ are the normalized left and right eigenfunctions of the tilted generator, we can write the eigenvalue $\Theta(\alpha,\beta)$ as  
\begin{equation}
\Theta(\alpha,\beta)=\int d\psi d\psi' \sum_i w_i(\psi,\psi')\, f_L(\psi)\left[e^{-\alpha_i(\psi,\psi')}f_R(\psi')-f_R(\psi)\right]+\int d\psi f_L(\psi)\left[\operatorname{grad}f_R(\psi)\cdot \mathcal{B}[\psi]-\beta(\psi)f_R(\psi)\right]\, .
\label{theta}
\end{equation}
In order to perform the maximization in (\ref{LegTransf}) we consider functional derivatives of $F(\alpha,\beta)$ with respect to the conjugated fields $\alpha,\beta$; we thus have ({\it cf.} Eqs.~\eqref{LegTransf}-\eqref{theta}):
\begin{equation}
\begin{split}
\frac{\delta F(\alpha,\beta)}{\delta\, \beta(\psi)}&=-\mu(\psi)+f_L(\psi)f_R(\psi)\, ,\\
\frac{\delta F(\alpha,\beta)}{\delta\,\alpha(\psi,\psi')}&=-q^i(\psi,\psi')+e^{-\alpha_i(\psi,\psi')}f_L(\psi)f_R(\psi')w_i(\psi,\psi')\, .
\end{split}
\end{equation}
For the special values $\alpha^*_i(\psi,\psi'),\, \beta^*(\psi)$, the above derivatives can be set to zero and one 
identifies
\begin{equation}
\begin{split}
\mu(\psi)=&f_L(\psi)f_R(\psi)\\
q^i(\psi,\psi')=&e^{-\alpha_i(\psi,\psi')}f_L(\psi)f_R(\psi')w_i(\psi,\psi')\, .
\end{split}
\end{equation}
Inverting the second of the above relations in order to explicitly write $\alpha_i(\psi,\psi')$, and substituting these new pieces of information in the functional $F(\alpha^*,\beta^*)$ we get
\begin{equation*}
\begin{split}
F(\alpha^*,\beta^*)=&-\sum_i\int d\psi d\psi' \left(q^i(\psi,\psi')-\mu(\psi)w_i(\psi,\psi')\right)-\int d\psi \frac{1}{f_R(\psi)}\operatorname{grad}f_R(\psi)\cdot \mathcal{B}[\psi]\mu(\psi)+\\
&+\sum_i\int d\psi d\psi' \, q^i(\psi,\psi')\log\left(\frac{q^i(\psi,\psi')}{f_L(\psi)f_R(\psi')w_i(\psi,\psi')}\right)\, .
\end{split}
\end{equation*}
We now decompose the integrand in the last term of the above equation in the following way
$$
q^i(\psi,\psi')\log\left(\frac{q^i(\psi,\psi')}{f_L(\psi)f_R(\psi')w_i(\psi,\psi')}\right)=q^i(\psi,\psi')\log\left(\frac{q^i(\psi,\psi')}{\mu(\psi)w_i(\psi,\psi')}\right)+q^i(\psi,\psi')\log\left(\frac{f_R(\psi)}{f_R(\psi')}\right)\, ,
$$
and we rewrite the level 2.5 functional $I_{2.5}[\mu,q]=F(\alpha^*,\beta^*)$ as 
\begin{equation}
\begin{split}
I_{2.5}[\mu,q]&=\int d\psi d\psi' \sum_i\Drel\left[q^i(\psi,\psi')|\mu(\psi)w_i(\psi,\psi')\right]+\\
&+\sum_i\int d\psi d\psi' q^i(\psi,\psi')\log\left(\frac{f_R(\psi)}{f_R(\psi')}\right)-\int d\psi \frac{1}{f_R(\psi)}\operatorname{grad}f_R(\psi)\cdot \mathcal{B}[\psi]\mu(\psi)\, ,
\label{funct2.5}
\end{split}
\end{equation}
with $\Drel$ being the relative entropy $\Drel[x|y]=x\log (x/y)-x+y$. In order to obtain the functional appearing in Eq. (\ratefn) in the main text, we shall now show that the term in the second line of the previous equation is zero when the continuity condition (\contin)
is satisfied.  

To this end, we notice that 
$$
\int d\psi d\psi' q^i(\psi,\psi')\log\left(\frac{f_R(\psi)}{f_R(\psi')}\right)=\int d\psi \log f_R(\psi)\left[\int d\psi'\left(q^i(\psi,\psi')-q^i(\psi',\psi)\right)\right]\, ,
$$
which, using the continuity condition, we rewrite as 
$$
\sum_i\int d\psi d\psi' q^i(\psi,\psi')\log\left(\frac{f_R(\psi)}{f_R(\psi')}\right)=-\int d\psi \log f_R (\psi)\operatorname{div}\left(\mathcal{B}[\psi]\mu(\psi)\right)\, .
$$
Finally, integrating by parts, we have 
$$
\sum_i\int d\psi d\psi' q^i(\psi,\psi')\log\left(\frac{f_R(\psi)}{f_R(\psi')}\right)=\int d\psi \frac{1}{f_R(\psi)}\operatorname{grad}f_R(\psi)\cdot \mathcal{B}[\psi]\mu(\psi)\, ;
$$
this shows that the second line of Eq.~\eqref{funct2.5} is zero and proves Eq.~(\ratefn) in the main text. To obtain Eq.~(\ratevar), it is sufficient to consider auxiliary processes with rates $w_i^A(\psi,\psi')$ such that the stationary measure $P^A_\infty(\psi)=\mu(\psi)$ and $P_\infty^Aw_i^A(\psi,\psi')=q^i(\psi,\psi')$. The large deviation function $I_{2.5}[\mu,q]$ will be then given by the optimal process minimizing the resulting functional. 

\subsection{Contraction from level 2.5 to level 1 LD for quantum systems}

This section shows how level-1 LDPs can be obtained by contraction from level 2.5, and yield the same results as are available via tilted Lindblad equations.  The general method is a standard one~[\citeHugo]: we minimise $I_{2.5}$ subject to constraints on (one or more) linear combinations of jump rates.  This is achieved by using a (vectorial) Lagrange multiplier $\vec\lambda$ to enforce the constraints.  The minimisation conditions for $I_{2.5}$ can be written as an eigenvalue problem, which is equivalent to finding the largest eigenvalue of a tilted Lindblad operator.

\subsubsection{Tilted operator approach}

As in the main text, let  ${Q}_\tau$ be vector whose elements $Q^i$ are the empirical rates of quantum jumps of type $i$, which are related to the number $N^i_\tau$ of such events as ${Q}_\tau^i=\frac{1}{\tau}N^i_\tau$. For long observation times $\tau$, this probability obeys a LD principle 
$$
\mathrm{Prob}[{Q}_\tau \approx Q]\approx {\rm e}^{-\tau\, I_1[Q]}\, ,
$$
From tilted operator techniques applied to full counting statistics~[\citeJuanpeFCS] it is known that such a LD function $I_1[Q]$ can be obtained by performing a Legendre transform 
\begin{equation}
I_1[Q]=\sup_{\vec{\lambda}}\left\{-Q\cdot \vec{\lambda}-\theta(\vec{\lambda})\right\}\, ,
\label{THEPHI}
\end{equation}
where $\vec{\lambda}$ is a vector of parameters conjugated to the various rates $Q^i$'s, and $\theta(\vec{\lambda})$ is the scaled cumulant generating function of the various time-integrated observables $Q^i$. This $\theta(\vec\lambda)$ is the largest real eigenvalue of the operator 
\begin{equation}
\mathcal{L}^\dagger_{\vec{\lambda}}[X]=i[H,X]+\sum_i \left({\rm e}^{-\lambda_i} J^\dagger_i X\, J_i -\frac{1}{2}\left[ X J^\dagger_i J_i +J^\dagger_i J_i X \right]\right)\, .
\label{TLO}
\end{equation}
where the $^\dagger$ on the left-hand-side indicates that this is the adjoint of a tilted version of the Lindblad operator in Eq.~(\lind).  If this eigenproblem can be solved then one may obtain the level-1 rate function as
\begin{equation}
I_1[Q]=-Q\cdot \vec{\lambda^*}-\theta(\vec{\lambda^*})\, .
\label{LDRF1}
\end{equation}
where $\lambda^*$ solves the maximisation problem in (\ref{THEPHI}), that is
\begin{equation}
-Q^i-\frac{\partial}{\partial\lambda_i}\theta(\vec{\lambda})=0\, .
\label{ACTLEVEL1}
\end{equation}



\subsubsection{Derivation by contraction}

We now show that Eq.~(\ref{LDRF1}) can be obtained by contraction from level-2.5, as
\begin{equation}
I_1[Q]=\inf_{\mu,q}I_{2.5}[\mu,q]\, ,
\label{equ:I1-contract}
\end{equation}
where the minimisation is subject to three constraints: (i) the auxiliary process should have jump rates $Q$, that is
\begin{equation}
\int d\psi d\psi'\, q^i(\psi,\psi')=Q^i\, , \qquad \forall i\, .
\label{fluck}
\end{equation}
and (ii) the empirical measure must be normalised as $\int{\rm d}\psi\, \mu(\psi)=1$; and (iii) the empirical measure and flux must obey the continuity condition

\begin{equation}
\operatorname{div} [\mathcal{B}[\psi]\mu(\psi)]=\sum_i\int d\psi' \left[q^i(\psi',\psi)-q^i(\psi,\psi')\right]\, ,
\label{CES}
\end{equation}
To make explicit the connection to an auxiliary process, we write the fluxes $q^i(\psi,\psi')$ in terms of modified jump rates $w_i^{\rm A}(\psi,\psi')$,
\begin{equation}
q^i(\psi,\psi')=\mu(\psi)\, w^{\rm A}_i(\psi,\psi')\, ;
\label{qaux}
\end{equation}
so that minimisation of $I_{2.5}$ over $q$ may be replaced by a minimisation over the jump rates of the modified process.

The constrained minimisation is implemented by introducing Lagrange multipliers. We thus define the extended functional 
\begin{equation}
\begin{split}
I[\mu,q,\lambda_0,\vec{\lambda},\Lambda)]=&\int d\psi d\psi'\, \mu(\psi)\sum_i\left[w^{\rm A}_i(\psi,\psi')\log\frac{w^{\rm A}_i(\psi,\psi')}{w_i(\psi,\psi')}-w^{\rm A}_i(\psi,\psi')+w_i(\psi,\psi')\right]
\\&+\lambda_0\left(\int d\psi \,\mu(\psi)-1\right)
+\sum_i\lambda_i\left(\int d\psi d\psi'\,\mu(\psi) w^{\rm A}_i(\psi,\psi')-Q^i\right)
\\
&+\int d\psi \, \Lambda(\psi)\left[\operatorname{div} [\mathcal{B}[\psi]\mu(\psi)]-\sum_i\int d\psi' \left(\mu(\psi')w^{\rm A}_i(\psi',\psi)-\mu(\psi) w^{\rm A}_i(\psi,\psi')\right)\right]\, .
\end{split}
\label{BigF}
\end{equation}
The Lagrange multiplier $\lambda_0$ takes care of the normalization of the empirical measure; also $\lambda_i$, with $i\neq0$, implement the constraints \eqref{fluck}, and the function $\Lambda(\psi)$ ensures \eqref{CES}. 
We now take a functional derivative with respect to 
$w_i^{\rm A}(\psi,\psi')$.  At the extremum we have
\begin{equation}
\begin{split}
0 = \frac{\delta I[\mu,q,\lambda_0,\vec{\lambda},\Lambda]}{\delta \,w^{\rm A}_i(\psi,\psi')}=\mu(\psi)\left[\log\frac{w^{\rm A}_i(\psi,\psi')}{w_i(\psi,\psi')}+\lambda_i-\Lambda(\psi')+\Lambda(\psi)\right]\, .
\label{Min2}
\end{split}
\end{equation}
Since $\mu>0$ we have
\begin{equation}
w^{\rm A}_i(\psi,\psi')=w_i(\psi,\psi'){\rm e}^{-\lambda_i}\frac{{\rm e}^{\Lambda(\psi')}}{{\rm e}^{\Lambda(\psi)}}\, ,
\label{C1}
\end{equation}
Next, consider the functional derivative of (\ref{BigF}) with respect to $\mu$:
\begin{equation}
\begin{split}
0 = \frac{\delta I[\mu,q,\lambda_0,\vec{\lambda},\Lambda]}{\delta \, \mu(\psi)}=&\int d\psi'\, \sum_i\left[w^{\rm A}_i(\psi,\psi')\log\frac{w^{\rm A}_i(\psi,\psi')}{w_i(\psi,\psi')}-w^{\rm A}_i(\psi,\psi')+w_i(\psi,\psi')\right]
\\&+ \lambda_0
+\sum_i\lambda_i\int d\psi' w^{\rm A}_i(\psi,\psi')
\\&+\frac{\delta}{\delta\,\mu(\psi)}\int d\psi \,\Lambda(\psi) \operatorname{div} [\mathcal{B}[\psi]\mu(\psi)]-\sum_i\int d\psi' \, \Lambda(\psi')w_i^{\rm A}(\psi,\psi')+\sum_i\int d\psi' \Lambda(\psi)w_i^{\rm A}(\psi,\psi')\, .
\label{Min1}
\end{split}
\ee
Using (\ref{C1}), this simplifies to
\begin{equation}
0 = \sum_i\int d\psi'w_i(\psi,\psi')\left[1-{\rm e}^{-\lambda_i}\frac{{\rm e}^{\Lambda(\psi')}}{{\rm e}^{\Lambda(\psi)}}\right]+\lambda_0+\frac{\delta}{\delta\, \mu(\psi)}\int d\psi \,\Lambda(\psi) \operatorname{div} [\mathcal{B}[\psi]\mu(\psi)]\, .
\label{EP}
\end{equation}
 Integrating by parts the divergence term, one gets
$$
\frac{\delta}{\delta\, \mu(\psi)}\int d\psi \, \Lambda(\psi) \operatorname{div} [\mathcal{B}[\psi]\mu(\psi)]=-\frac{\delta}{\delta\,\mu(\psi)}\int d\psi \, \operatorname{grad}\Lambda(\psi)\cdot \mathcal{B}[\psi]\mu(\psi)=-\operatorname{grad}\Lambda(\psi)\cdot \mathcal{B}[\psi]\, ,
$$
Hence one may write \eqref{EP} as an eigenvalue problem for the positive function ${\rm e}^{\Lambda(\psi)}$
\begin{equation}
\mathcal{W}_{\vec{\lambda}}\left[{\rm e}^{\Lambda(\psi)}\right]=\lambda_0\, {\rm e}^{\Lambda(\psi)}\, ,
\label{EP2}
\end{equation}
where the generator $\mathcal{W}_{\vec{\lambda}}$ is the integro-differential operator given by 
$$
\mathcal{W}_{\vec{\lambda}}[f(\psi)]=\sum_i\int d\psi'\, w_i(\psi,\psi')\left[{\rm e}^{-\lambda_i}f(\psi')-f(\psi)\right]+\operatorname{grad} f(\psi)\cdot \mathcal{B}[\psi]\, .
$$
This may be viewed as a tilted version of the UQME generator $\cal W$ defined in (\ref{equ:Wf}).

Assuming both the initial process and the above generator $\mathcal{W}_{\vec{\lambda}}$ to possess a unique largest real eigenvalue, we can invoke Perron-Frobenius theorem: this means that if we find a positive eigenfunction of the operator $\mathcal{W}_{\vec{\lambda}}$, then this eigenfunction is unique and associated to the largest real eigenvalue of the operator.  
Assume (as will be verified below) that this eigenfunction can be expressed as
\be
{\rm e}^{\Lambda(\psi)}= \Tr(\psi\, \ell),
\label{equ:Lambdasol}
\ee
where $\ell=\ell^\dagger$ is a positive matrix. We then substitute this into equation \eqref{EP2}. After some calculations, one finds 
\be
\Tr\left(\psi\,  \mathcal{L}^\dagger_{\vec{\lambda}}[\ell]\right)=\lambda_0\Tr\left(\psi\,  \ell\right)\, ,
\label{equ:L25-lind}
\ee
where $\mathcal{L}^\dagger_{\vec{\lambda}}$ is the same operator as in \eqref{TLO}.  Moreover, this has to be true for all $\psi$ so
\be
\mathcal{L}^\dagger_{\vec{\lambda}}[\ell] = \lambda_0  \ell
\label{equ:lind-tilt-ell}
\ee
and we recognise $\ell=\ell_{\vec\lambda}$ as the dominant eigenmatrix of the tilted Lindblad operator from (\ref{TLO}).
That is, the $\Lambda$ that minimises (\ref{BigF}) is related to the tilted Lindbladian via (\ref{equ:Lambdasol}).
Also $\lambda_0 = \theta(\vec\lambda)$, which is the eigenvalue that appears in (\ref{THEPHI}).
This establishes the first connection between the level-2.5 approach and the tilted Lindbladian.

Thus far, we have considered variations of (\ref{BigF}) with respect to $w^A$ (or equivalently $q$) and $\mu$.  From (\ref{equ:Lambdasol},\ref{equ:lind-tilt-ell},\ref{C1}) we have enough conditions to fix $w^A$, $\Lambda$ and $\lambda_0$ to specific values $((w^A)^*,\Lambda^*,\lambda_0^*)$ which all depend on the $\lambda_i$.  Note in particular that $\lambda_0^*=\theta(\vec\lambda)$. Next, consider variations of  (\ref{BigF}) with respect to $\Lambda,\lambda_0$: these enforce the continuity condition and normalisation of $\mu$. Hence, $\mu$ must be the (normalised) invariant measure for the auxiliary process with rates $(w^A)^*$, which we denote by $\mu^*$.  For compactness of notation we omit the stars on $w^A$: also define $q^*(\psi,\psi')=\mu^*(\psi) w^A(\psi,\psi')$.
Substituting into (\ref{BigF}), we obtain
\begin{equation}
\begin{split}
I[\mu^*,q^*,\lambda_0^*,\vec{\lambda},\Lambda^*]=&\int d\psi d\psi'\, \mu^*(\psi)\sum_i\left[w^{\rm A}_i(\psi,\psi') \left[-\lambda_i+\Lambda^*(\psi')-\Lambda^*(\psi)\right]-w^{\rm A}_i(\psi,\psi')+w_i(\psi,\psi')\right] \\
& + \sum_i\lambda_i\left(\int d\psi d\psi'\,\mu^*(\psi) w^{\rm A}_i(\psi,\psi')-Q^i\right)
\label{equ:almost-there}
\end{split}
\end{equation}
Note carefully that all the starred quantities (and $w^A$) depend on $\vec\lambda$ since we have extremised (\ref{BigF}) at fixed $\vec\lambda$.

At this point we notice that consistency with (\ref{equ:I1-contract}) requires that extremising (\ref{equ:almost-there}) over $\vec\lambda$ recovers $I_1(Q)$.  For this it is sufficient that $I[\mu^*,q^*,\lambda_0^*,\vec{\lambda},\Lambda^*]$ coincides with the quantity $-Q\cdot\vec\lambda - \theta(\vec\lambda)$ which appears in (\ref{THEPHI}).  We will now verify this by using properties of the starred objects.
First, use (\ref{EP2}) to write
$$
{\rm e}^{-\Lambda^*(\psi)} \sum_i\int d\psi'\, w_i(\psi,\psi')\left[{\rm e}^{-\lambda_i}{\rm e}^{\Lambda^*(\psi')}-{\rm e}^{\Lambda^*(\psi)}\right]+ {\rm e}^{-\Lambda^*(\psi)}  \mathcal{B}[\psi] \cdot \operatorname{grad}{\rm e}^{\Lambda^*(\psi)}  = \lambda_0^*
$$
Multiply by $\mu^*(\psi)$, integrate over $\psi$, and use (\ref{C1}): one finds
$$
\sum_i \int d\psi\,d\psi'\, \mu^*(\psi) \left[ w^A_i(\psi,\psi') - w_i(\psi,\psi')  \right] + \int d\psi\, \mu^*(\psi) {\cal B}[\psi] \cdot \operatorname{grad} \Lambda^*(\psi) = \theta(\vec\lambda)
$$
where we also used that $\lambda_0^* = \theta(\vec\lambda)$.
Integrating by parts in the last term, and applying the continuity condition (\ref{CES}), one finds (after exchanging the integration variables $(\psi,\psi')$ in one term) that
\be
\sum_i \int d\psi\,d\psi'\, \left\{ \mu^*(\psi) \left[ w^A_i(\psi,\psi') - w_i(\psi,\psi') + w^A_i(\psi,\psi') [ \Lambda^*(\psi) - \Lambda^*(\psi') \right]
 \right\}
 = \theta(\vec\lambda)
 \label{equ:what-we-want}
\ee

Finally, we return to (\ref{equ:almost-there}) and collect terms: we obtain
$$
I[\mu^*,q^*,\lambda_0^*,\vec{\lambda},\Lambda^*]=\int d\psi d\psi'\, \mu^*(\psi)\sum_i\left[w^{\rm A}_i(\psi,\psi')^* 
\left[\Lambda^*(\psi')-\Lambda^*(\psi)\right]-w^{\rm A}_i(\psi,\psi')^*+w_i(\psi,\psi')\right] - Q \cdot \vec\lambda
$$
Comparing with (\ref{equ:what-we-want}), we see that indeed
\be
I[\mu^*,q^*,\lambda_0^*,\vec{\lambda},\Lambda^*] = - Q \cdot \vec\lambda -\theta(\vec\lambda) 
\ee
so that taking the extremum with respect to $\vec\lambda$ recovers (\ref{THEPHI}), as required.   That is, the level-1 LDP can be obtained by contraction from level-2.5.

\subsection{Connection between auxiliary process and quantum Doob transformation}

This section shows that quantum Doob transform of [\citeDoob] is related to the optimally-controlled auxiliary process discussed in this work.
The central idea is that the dynamical evolution of $\psi_t$ in the auxiliary process (with rates $w^A$) cannot be described by a Belavkin equation in the form (\belav).  (The jump operators $J_i$ in (\belav) control the end points of the jumps, and the rates of the jumps, and they determine the anti-Hermitian part of $H_{\rm eff}$.  In the auxiliary process then one may write a stochastic differential equation for $\psi$ but these different aspects of the dynamics would be controlled by different operators.)  However, one may construct a matrix $\Psi_t$ that evolves according to the Belavkin equation, with the jump operators $J_i$ and the Hamiltonian $H$ replaced by $\tilde J_i$ and $\tilde H$ respectively.  This matrix is an unravelling of the quantum Doob process.
We take
\be
\Psi_t=
\frac{\ell_{\vec{\lambda}}^{1/2}\psi_t \ell_{\vec{\lambda}}^{1/2}}
{\Tr\left(\ell_{\vec{\lambda}}\psi_t\right)}\, .
\label{equ:def-Psi}
\ee


Given a jump $\psi_t\to \frac{\mathcal{J}_i[\psi_t]}{\Tr\left(\mathcal{J}_i[\psi_t]\right)}$ in the auxiliary process, one sees that the corresponding jump for $\Psi_t$ is
$$
\Psi_t\to \frac{\ell_{\vec{\lambda}}^{1/2}\mathcal{J}_i[\ell_{\vec{\lambda}}^{-1/2}\Psi_t \ell_{\vec{\lambda}}^{-1/2}]\ell_{\vec{\lambda}}^{1/2}}
{\Tr\left(\ell_{\vec{\lambda}}\mathcal{J}_i[\ell_{\vec{\lambda}}^{-1/2}\Psi_t \ell_{\vec{\lambda}}^{-1/2}]\right)}\, .
$$
Recalling the definition $\cal J[\psi] = J\psi J^\dag$, this motivates us to introduce a new set of jump operators 
$
 \tilde{J}_i \propto \ell_{\vec{\lambda}}^{1/2}\,J_i\, \ell_{\vec{\lambda}}^{-1/2}
$
and similarly $\tilde{\mathcal{J}}_i[\Psi] =  \tilde{J}_i \Psi  \tilde{J}_i^\dag$
so that a jump $\psi\to \frac{\mathcal{J}_i[\psi_t]}{\Tr\left(\mathcal{J}_i[\psi_t]\right)}$ results in a jump $\Psi\to \frac{\tilde{\mathcal{J}}_i[\Psi_t]}{\Tr\left(\tilde{\mathcal{J}}_i[\Psi_t]\right)}$.  
From Eq(\wrates) of the main text, one sees that $\Psi_t$ can be obtained by unravelling a Lindblad process only if the rate for such jumps is $\Tr(\tilde{\mathcal{J}}_i[\Psi_t])$.  To verify this we use (\ref{C1},\ref{equ:Lambdasol}) and (\wrates) to see that the jump rate out of $\psi$ is 
$$
\int w^A_i(\psi,\psi') {\rm d}\psi' = \frac{{\rm e}^{-\lambda_i}\Tr\left(\ell_{\vec{\lambda}}\mathcal{J}_i[\psi_t]\right)}{\Tr\left(\ell_{\vec{\lambda}}\psi_t\right)}
= \Tr(\tilde{\mathcal{J}}_i[\Psi_t]) $$
where we have now fixed the constant of proportionality in $\tilde{J}_i$:
\be
 \tilde{J}_i = {\rm e}^{-\lambda_i/2}\, \ell_{\vec{\lambda}}^{1/2}\,J_i\, \ell_{\vec{\lambda}}^{-1/2} .
\label{equ:tilde-J-const}
\ee


Finally, we must show that the evolution of $\Psi$ between jumps follows a Belavkin equation in which the anti-Hermitian part of $H_{\rm eff}$ coincides with $\sum_i \tilde{J}_i^\dag \tilde{J}_i$.  For this deterministic evolution we have
$$
\dot{\Psi}_t=\frac{\ell_{\vec{\lambda}}^{1/2}\,\dot{\psi}_t \,\ell_{\vec{\lambda}}^{1/2}}{\Tr\left(\ell_{\vec{\lambda}}^{1/2}\,\psi_t \,\ell_{\vec{\lambda}}^{1/2}\right)}-\frac{\ell_{\vec{\lambda}}^{1/2}\,{\psi}_t\, \ell_{\vec{\lambda}}^{1/2}\Tr\left(\ell_{\vec{\lambda}}^{1/2}\,\dot{\psi}_t\, \ell_{\vec{\lambda}}^{1/2}\right)}{\Tr\left(\ell_{\vec{\lambda}}^{1/2}\,\psi_t \,\ell_{\vec{\lambda}}^{1/2}\right)^2}\, ;
$$
where the dot indicates a time derivative.
To shorten notation write $G=-iH_{\rm eff}$, so that (\belav) yields (for time periods between jumps)
$$
\dot{\psi}_t=G\psi_t +\psi_t G^\dagger -\psi_t \Tr\left(G\psi_t +\psi_t G^\dagger\right)\, .
$$
After some algebra we obtain
$$
\dot{\Psi}_t=\frac{\ell_{\vec{\lambda}}^{1/2}(G\psi_t +\psi_t G^\dagger) \ell_{\vec{\lambda}}^{1/2}}{\Tr\left(\ell_{\vec{\lambda}}^{1/2}\,\psi_t \,\ell_{\vec{\lambda}}^{1/2}\right)}-\frac{\ell_{\vec{\lambda}}^{1/2}\,\psi_t \, \ell_{\vec{\lambda}}^{1/2} }{\Tr\left(\ell_{\vec{\lambda}}^{1/2}\,\psi_t \,\ell_{\vec{\lambda}}^{1/2}\right)^2}\Tr\left(\ell_{\vec{\lambda}}^{1/2}(G\psi_t +\psi_t G^\dagger )\ell_{\vec{\lambda}}^{1/2} \right)\, ,
$$
which considering the definition of $\Psi_t$ and introducing the operator $\hat{G}=\ell_{\vec{\lambda}}^{1/2}\,G\,\ell_{\vec{\lambda}}^{-1/2}$ can be written as
\be
\dot{\Psi}_t=\hat{G}\Psi_t +\Psi_t\hat{G}^\dagger -\Psi_t \Tr\left(\Psi_t(\hat{G}+ \hat{G}^\dagger )\right)\, .
\label{equ:Psi-det}
\ee
showing that the time evolution of $\Psi$ between jumps is analogous to that of $\psi$, but with $G$ replaced by $\hat G$.
Now observe that 
$$
\hat{G}+\hat{G}^\dagger
=\ell_{\vec{\lambda}}^{-1/2}\Big(\ell_{\vec{\lambda}} G+G^\dagger \ell_{\vec{\lambda}}\Big)\ell_{\vec{\lambda}}^{-1/2}\, .
$$
Adding and subtracting the term $\sum_i {\rm e}^{-\lambda_i} \mathcal{J}^\dagger_i[\ell_{\vec{\lambda}}]$ inside the parenthesis
and using (\ref{TLO}), one gets
$$
\hat{G}+\hat{G}^\dagger=\ell_{\vec{\lambda}}^{-1/2}\mathcal{L}^{\dagger}_{\vec{\lambda}}[\ell_{\vec{\lambda}}]\ell_{\vec{\lambda}}^{-1/2}-\ell_{\vec{\lambda}}^{-1/2}\Big(\sum_i {\rm e}^{-\lambda_i} \mathcal{J}^\dagger_i[\ell_{\vec{\lambda}}] \Big)\ell_{\vec{\lambda}}^{-1/2}\, , 
$$
Now recall that $\ell$ is an eigenmatrix: it obeys (\ref{equ:lind-tilt-ell}) so that
$$
\hat{G}+\hat{G}^\dagger=\theta(\vec{\lambda})-\ell_{\vec{\lambda}}^{-1/2}\Big(\sum_i {\rm e}^{-\lambda_i} \mathcal{J}^\dagger_i[\ell_{\vec{\lambda}}] \Big)\ell_{\vec{\lambda}}^{-1/2}\, .
$$
Introducing the new Hamiltonian $\tilde{H}=i(\hat{G}-\hat{G}^\dagger)/2$,  the deterministic evolution (\ref{equ:Psi-det}) reduces to 
$$
\dot{\Psi}_t= -i\tilde{H}_{\rm eff}\Psi_t +i \Psi_t \tilde{H}_{\rm eff}^\dagger -\Psi_t \Tr\left(-i\tilde{H}_{\rm eff}\Psi_t +i \Psi_t \tilde{H}_{\rm eff}^\dagger\right) \, .
$$
 where $\tilde{H}_{\rm eff}=\tilde{H}-\frac{i}{2}\sum_i \tilde{J}_i^\dagger \tilde{J}_i$, which is also equal to $i[\hat G-\theta(\vec\lambda)/2]$.
  
Together with the fact that also the various quantum jumps are controlled by the new operators $\tilde{J}_i$, 
this establishes that $\Psi$ obeys a Belavkin equation similar to (\belav), with Hamiltonian $\tilde H$ and jump operators $\tilde J_i$.  Taking the average of this $\dot\Psi$, we find that the dynamics of the mixed density matrix is controlled by the new Lindblad operator
\be
\tilde{\mathcal{L}}[{\rho}_t]=-i[\tilde{H},{\rho}_t]+\sum_i\left(\tilde{J}_i {\rho}_t \tilde{J}_i^\dagger -\frac{1}{2}\left[ {\rho}_t \tilde{J}_i^\dagger \tilde{J}_i +  \tilde{J}_i^\dagger \tilde{J}_i {\rho}_t \right] \right)\, .
\label{equ:lindblad-doob}
\ee
This average dynamics coincides with the Doob dynamics studied in [\citeDoob].



\section{Applications}

\subsection{Quantum reset processes and auxiliary rates}


This section derives some properties of quantum reset processes that are used in the main text.
We consider quantum processes obeying (\belav)
where we assume that all jump operators project onto pure states $J_i\psi J^\dagger_i\propto \varphi_i$. 
As usual for (\belav), dynamical trajectories consist of deterministic segments with random jumps.  The special feature of quantum reset processes is that the probability to jump at time $t$ depends only on the type of the previous jump and the time at which this jump took place.


After each jump, therefore, the system is in one of the states $\varphi_i$.  Suppose (without loss of generality) that this jump takes place at time zero.
Then the probability that system does not jump between time zero and time $t$ is
$$
s_i(t)=\Tr\left({ e}^{-iH_{\rm eff}t}\varphi_i{e}^{iH^\dagger_{\rm eff}t}\right)\, 
$$
where $H_{\rm eff}=H-i/2\sum_iJ^\dagger_i J_i$, as in the main text.
Let $p_{ij}(t){\rm dt}$ be the probability that the first jump after time zero takes place between time $t$ and $t+{\rm d}t$.  We have
$$
p_{ij}(t)=\Tr\left(J_j\, {e}^{-iH_{\rm eff}t}\varphi_i{e}^{iH^\dagger_{\rm eff}t}\, J_j^\dagger\right)\, .
$$
These probabilities are normalised such that $\sum_{j} \int_0^\infty p_{ij}(t) {\rm d}t=1$, and one has 
\be
{s}_i(t) = \sum_j \int_t^\infty p_{ij}(t') {\rm d}t'
\label{equ:sp}
\ee
with $s_i(0)=1$ as required.  Equ.(\ref{equ:sp}) is most easily verified by differentiating both sides with respect to $t$ and using the definitions of $s$ and $p$.
Also define 
\be
R_{ij} = \int_0^\infty p_{ij}(t) {\rm d}t
\label{equ:def-R}
\ee
which is the probability that the first jump after time zero will be into state $\varphi_j$, given that the last jump (at time zero) was of type $i$.
The steady-state probability distribution (invariant measure) of the process is determined by the jump rates into $\varphi_i$, the deterministic evolution away from this starting point, and the survival probability.  Specifically, 
\be
P_\infty(\psi) = \sum_i {c}_i\int_0^\infty dt\, {s}_i(t)\, \delta(\psi-\varphi_{i t} )
\label{equ:Pinf}
\ee
where $c_i$ is the (so far undetermined) total rate of jumps into $\varphi_i$ (in the steady state) and
\be
\varphi_{i t}=\frac{{ e}^{-iH_{\rm eff}t}\varphi_i{ e}^{iH^\dagger_{\rm eff}t}}{\Tr\left({ e}^{-iH_{\rm eff}t}\varphi_i{ e}^{iH^\dagger_{\rm eff}t}\right)}\, .
\label{equ:varphi}
\ee
The steady state rate of jumps into into $\varphi$ can also be obtained as a steady-state average of the jump rate: $c_i = \int P_\infty(\psi) \Tr({\cal J}_i\psi) {\rm d}\psi$.  Using the definition of $p_{ij}(t)$ and $R_{ij}$ this implies
\be
{c}_i =\sum_j {c}_j {R}_{ji}\, 
\label{equ:c-bal}
\ee
Noting that $\sum_j R_{ij}=1$, one may consider a discrete-time Markov chain in which the transition probabilities are $R_{ij}$, in which case $c_i$ is proportional to the steady state occupancy of state $i$.  However, the $c_i$ are rates (and not probabilities), their normalisation is obtained by insisting that $P_\infty$ in (\ref{equ:Pinf}) is a normalised distribution over matrices $\psi$.  With this choice it may be verified that (\ref{equ:Pinf}) is indeed a steady-state solution of the UQME (\uqme).


The full statistics of quantum jumps can be characterised in terms of $s_i(t)$ and $p_{ij}(t)$.
To analyse large deviations of quantum jumps in these processes, we compare them with auxiliary processes that are formulated directly in terms of the waiting times between jumps.  That is, we consider an auxiliary (modified) process that is characterised by the functions
%
$\hat{p}_{ij}(t)$ and 
\be 
\hat{s}_i(t) = \sum \int_t^\infty \hat p_{ij}(t') {\rm d}t' \; .
\label{equ:sphat}
\ee
Repeating the analysis above, one arrives at corresponding definitions of $\hat{R}_{ij}, \hat{c}_i$.  The invariant measure associated with this modified process will be denoted by $\mu(\psi)$, which plays the role of $\hat{P}_\infty(\psi)$: that is
\begin{equation}
{\mu}(\psi)=\sum_i \hat{c}_i\int_0^\infty dt\, \hat{s}_i(t)\, \delta(\psi-\varphi_{it})\, ,
\label{TiltEmp}
\end{equation}

We may also derive the auxiliary rates $w^A$  that correspond to this model.  Using  $q_i(\psi,\psi') = \mu(\psi) w^A_i(\psi,\psi')$ and (\wA) in (\contin), one obtains
%
%
%
%
%

$$
\operatorname{div} \left[\mathcal{B}[\psi]{\mu}(\psi)\right]
-\sum_i\int d\psi'\, \left[ \mu(\psi') A_i(\psi') w_i(\psi',\psi)- \mu(\psi) A_i(\psi) w_i(\psi,\psi')\right]=0\, .
$$
The function $\cal B$ is fixed by the model and $\mu$ is given by (\ref{TiltEmp}) -- we seek a suitable $A_i$, which needs to be defined only within the support of $\mu$.  That is, it is sufficient to specify $A_j(\varphi_{i t})$ for all $i,j,t$.  We will show that the solution is
\be
A_i(\varphi_{j t})=\frac{\hat{p}_{ji}(t)}{\hat{s}_j(t)}\frac{1}{\Tr\left(J_i\varphi_{jt}J_i^\dagger\right)}\, .
\label{equ:A-renew}
\ee

To verify this, it is convenient to introduce a general (scalar) function $f$ and to write
\begin{equation}
\int d\psi \, f(\psi) \Bigg(\operatorname{div} \left[\mathcal{B}[\psi]{\mu}(\psi)\right]-\sum_i\int d\psi'\, 
\left[ \mu(\psi') A_i(\psi') w_i(\psi',\psi)- \mu(\psi) A_i(\psi) w_i(\psi,\psi')\right]
\Bigg)=0\, .
\label{CESB}
\end{equation}
which must hold for all functions $f$.
We now proceed by simplifying term by term the different pieces of the above relation.
Using integration by parts and (\ref{TiltEmp}) we find
$$
\int d\psi f(\psi)\operatorname{div}\left[\mathcal{B}[\psi]{\mu}(\psi)\right]=-\int d\psi\operatorname{grad} f(\psi)\cdot \mathcal{B}[\psi]{\mu}(\psi)=-\sum_i \hat{c}_i\int_0^\infty dt \, \hat{s}_i(t)\, \operatorname{grad} f(\varphi_{it})\cdot \mathcal{B}[\varphi_{i t}]; 
$$
Using the chain rule of differentiation and then (\ref{equ:varphi}), one has $\frac{d}{dt}f(\varphi_{i t})=\operatorname{grad} f(\varphi_{i t})\cdot\dot\varphi_{i t}=\operatorname{grad} f(\varphi_{i t})\cdot \mathcal{B}[\varphi_{i t}]$.  Hence 
$$
\int d\psi f(\psi)\operatorname{div}\left[\mathcal{B}[\psi]{\mu}(\psi)\right]=-\sum_i \hat{c}_i\int_0^\infty dt\, \hat{s}_i(t)\, \frac{d}{dt} f(\varphi_{i t}). 
$$
Integrating again by parts we obtain 
\be
\int d\psi f(\psi)\operatorname{div}\left[\mathcal{B}[\psi]{\mu}(\psi)\right]=\sum_i \hat{c}_i f(\varphi_i)+\sum_i \hat{c}_i \int_0^\infty dt\, f(\varphi_{i t})\, \frac{d}{dt}\hat{s}_i(t). 
\label{equ:c-term1}
\ee
For the other terms in (\ref{CESB}) we exchange integration variables $\psi,\psi'$ and obtain
\begin{align}
 \sum_i  \int d\psi \, d\psi'\, [ f(\psi) - f(\psi') ]
\mu(\psi') A_i(\psi') w_i(\psi',\psi) 
& = \sum_{ij}  \hat{c}_j  \int_0^\infty {\rm d}t\,  \hat{s}_j(t) A_i(\varphi_{j t}) \Tr(J_i \varphi_{j t}  J_i^\dag) [f(\varphi_i) - f(\varphi_{j t}) ]
\nonumber \\ 
& = \sum_{j}   \left[ \sum_i \int_0^\infty dt \,  \hat{c}_j  \hat{p}_{ji}(t) f(\varphi_i) + \int_0^\infty {\rm d}t\,  f(\varphi_{j t}) \frac{d }{dt} \hat{s}_j(t) \right]\
\nonumber \\ 
& =  \sum_i \hat{c}_i f(\varphi_i) + \sum_{j}   \int_0^\infty {\rm d}t\,  f(\varphi_{j t}) \frac{d }{dt} \hat{s}_j(t)
\end{align}
The first equality uses the definition $w_i(\psi,\psi') = \Tr(J_i \psi  J_i^\dag)\delta(\psi'-\varphi_i)$ and (\ref{TiltEmp});
the second uses (\ref{equ:A-renew}) with (\ref{equ:sphat}) and $\hat{s}_i(0)=1$; and the third uses 
(\ref{equ:def-R}) applied to the auxiliary process (ie with $R\to\hat{R}$ and $c\to\hat{c}$) and (\ref{equ:c-bal}.
The final expression coincides with the right-hand-side of (\ref{equ:c-term1}), so one sees that the continuity condition (\ref{CESB}) is indeed satisfied for the modified rates in (\ref{equ:A-renew}).

%

To summarise, the modified rates of (\ref{equ:A-renew}) lead to a quantum reset process with the jump probabilities $\hat{p}_{ij}(t)$.
The steady-state distribution of this modified process is given by (\ref{TiltEmp}), where the $\hat{c}_i$ are obtained (up to an overall normalisation) by solving the balance condition (\ref{equ:c-bal}) for the modified process (ie with the replacements $c\to\hat{c}$ and $R\to\hat{R}$).

\subsection{Bounds for the level 1 LD function}
In the previous section we found the empirical measure and the jump rates for an auxiliary quantum reset processes satisfying the continuity equation. This information can now be used to provide a bound to the actual LD rate function for the jump activities. 

Consider the LD rate function $I_1[Q]$ for a specific vector $Q$. The first thing to do is to constrain the auxiliary process to produce exactly the activities $Q$. This can be done choosing appropriate probabilities $\hat{p}_{ij}(t)$ such that 
$$
Q_j=\int d\psi d\psi' \, q^j(\psi,\psi')=\int d\psi  d\psi' \mu(\psi) \, A_i(\psi)\Tr\left(J_j\psi J^\dagger_j\right)\delta(\psi'-\varphi_j)=\sum_i\hat{c}_i \hat{R}_{ij}=\hat{c}_j\, , 
$$
where the last equality follows from the discussion at the end of the previous section.

Then, we can substitute the details of the auxiliary process in the level $2.5$ functional thus obtaining 
$$
I_1[Q]\le I_{2.5}[\mu,q]= \sum_{i,j}Q_i\int_0^\infty dt\left[ \hat{p}_{ij}(t)\log \left[\frac{\hat{p}_{ij}(t)}{p_{ij}(t)}\frac{s_i(t)}{\hat{s}_i(t)}\right]-\hat{s}_i(t)\left(\frac{\hat{p}_{ij}(t)}{\hat{s}_i(t)}-\frac{p_{ij}(t)}{s_i(t)}\right)\right]\, ,
$$
where $s_i(t),p_{ij}(t)$ are the probabilities of the original reset process. Now we focus on the second term appearing in the above integral:
$$
\sum_j\int_0^\infty dt\, \hat{s}_i(t)\left(\frac{\hat{p}_{ij}(t)}{\hat{s}_i(t)}-\frac{p_{ij}(t)}{s_i(t)}\right)=\int_0^\infty dt\, \hat{s}_i(t)\left(-\frac{\frac{d\hat{s}
_i(t)}{dt}}{\hat{s}_i(t)}+\frac{\frac{d{s}
_i(t)}{dt}}{d{s}_i(t)}\right)=\int_0^\infty dt\, \hat{s}_i(t)\left(-\frac{d}{dt}\log\hat{s}_i(t)+\frac{d}{dt}\log s_i(t)\right)\, ,
$$
where, for the first equality, we have used the relation \eqref{equ:sphat}. Integrating by parts, one then obtains 
$$
\sum_j\int_0^\infty dt\, \hat{s}_i(t)\left(\frac{\hat{p}_{ij}(t)}{\hat{s}_i(t)}-\frac{p_{ij}(t)}{s_i(t)}\right)=-\sum_{j}\int_0^\infty dt\, \hat{p}_{ij}(t)\log \frac{\hat{s}_i(t)}{s_i(t)}\, ,
$$
which substituted in the functional provide the expression 
\begin{equation}
I_1[Q]\le \sum_{i,j} Q_i\int_0^\infty dt\, \hat{p}
_{ij}(t)\log\frac{\hat{p}_{ij}(t)}{p_{ij}(t)}\, .
\label{CostModProbab}
\end{equation}

\subsection{Thermodynamic uncertainty relations for the jump activities}
In order to obtain Eq.~($13$) in the main text we proceed as follows. We consider modified probabilities $\hat{p}_{ij}(t)$ given by 
$$
\hat{p}_{ij}(t)=v_{ij}e^{-u_{ij} t}p_{ij}(t);
$$
the first requirement that we want to satisfy is that $\hat{R}_{ij}=R_{ij}$, and we shall consider this up to second order in the quantities $u_{ij}$ which we take to be small. We thus have
$$
\hat{R}_{ij}=\int_0^\infty dt \, \hat{p}_{ij}(t)\approx v_{ij}R_{ij}\left(1-u_{ij}\langle t\rangle_{ij}+\frac{1}{2}u^2_{ij}\langle t^2\rangle_{ij} \right)\, ,
$$
where we have $\langle t\rangle_{ij}=R_{ij}^{-1}\int_0^\infty dt\,  t\, p_{ij}(t)$ and $\langle t^2\rangle_{ij}=R_{ij}^{-1}\int_0^\infty dt\,  t^2\, p_{ij}(t)$. To have this quantity equal to $R_{ij}$ up to second order in $u_{ij}$ we set 
$$
v_{ij}\approx 1+u_{ij}\langle t\rangle_{ij}+\left(\langle t\rangle_{ij}^2-\frac{1}{2}\langle t^2\rangle_{ij}\right)\, .
$$
Then we introduce the probabilities $\hat{p}_{ij}$ into the functional on the right hand side of \eqref{CostModProbab} and expanding up to second order in $u_{ij}$ we have 
\begin{equation}
\sum_{i,j} Q_i\int_0^\infty dt\, \hat{p}
_{ij}(t)\log\frac{\hat{p}_{ij}(t)}{p_{ij}(t)}\approx \sum_{i,j} \bar{Q}_i \frac{R_{ij}u_{ij}^2}{2}\sigma_{ij}^2\, ,
\label{TBound}
\end{equation}
where $\sigma_{ij}^2=R_{ij}^{-1}\int_0^\infty dt\, (t-\tau_{ij}^2)p_{ij}(t)$ and $\tau_{ij}=\langle t\rangle_{ij}$ as in the main text; also, $\bar Q_i=c_i$ is the (average) rate of jumps for the original process.
Notice that the jump rates $Q_i$ are taken to be the averages since on the right hand side of the above equation the term is already of second order in the $u_{ij}$'s.

We then consider the other requirement that we want, namely that the average jump times $\hat{\tau}_{ij}$ for the new probabilities are uniformly rescaled $\hat{\tau}_{ij}=\tau_{ij}/a$, with $a=1+\delta$ and $\delta$ being a small number. This gives 
$$
\hat{\tau}_{ij}=R_{ij}^{-1}\int_0^\infty dt \, t\, \hat{p}_{ij}\approx \tau_{ij}\left(1-u_{ij}\frac{\sigma_{ij}^2}{\tau_{ij}}\right)=\tau_{ij}(1-\delta);
$$
notice that in this case we only need to expand up to first order in the $u_{ij}$'s since these already appear up to second order in the right hand side of \eqref{TBound}. The above equaiton thus provides a relation for the $u_{ij}$ that must be $u_{ij}=\delta\frac{\tau_{ij}}{\sigma_{ij}^2}$. 

Since also the jumps $Q$ in this infinitesimaly modified process are rescaled by $Q=a\bar{Q}$, we have the relation 
$$
I_1[(1+\delta)\bar{Q}]\le \frac{1}{2}\chi \delta^2+O(\delta^3)
$$
with 
$$
\chi=\sum_{ij}\bar{Q}_iR_{ij}\frac{\tau_{ij}^2}{\sigma_{ij}^2}\, .
$$

\subsection{Bounds for the example system}

The system we consider here is made of a particle that can jump both coherently and incoherently between two sites. The state where the particle is in the first site is $\varphi_{\rm L} =|10\rangle\langle 10|$ while when the particle is in the second site we describe it through the state $\varphi_{\rm R}=|01\rangle\langle 01|$. The Hamiltonian is given by $H=\Omega \left(|10\rangle\langle 01|+|01\rangle \langle 10|\right)$, and jump operators are $J_{\rm L}=\sqrt{\gamma}|10\rangle \langle 01|$ and $J_{\rm R}=\sqrt{\gamma}|01\rangle\langle 10|$. The two reset states are thus $\varphi_{\rm L}, \varphi_{\rm R}$, which are separable states. 

We set here $\gamma=1$. As defined in the main text, quantum jumps in the system are characterized by the probabilities $p_{ij}(t)$, which are given in this specific case by:
\begin{equation}
p_{\rm L L}(t)=p_{\rm RR}(t)=e^{-t}\sin^2(\Omega\, t)\, , \qquad p_{\rm LR}(t)=p_{\rm RL}(t)=e^{-t}\cos^2(\Omega\, t)\, .
\end{equation}
In this case the two survival probabilities coincide and are exponential $s_{\rm L}(t)=s_{\rm R}(t)=e^{-t}$.

To find bounds for activity and entanglement we introduce the modified probabilities $\hat{p}_{ij}(t)=v_{ij}p_{ij}(t)e^{-u\, t}$ and we impose that 
$$
\int_0^\infty \hat{p}_{ij}(t)=\int_0^\infty {p}_{ij}(t)\, ,
$$
leaving $u$ as an independent variable and thus fixing the $v_{ij}$. Taking this into account we can write these probabilities (to give compact formulae we now fix $\Omega=1/2$) as 
\begin{equation}
\hat{p}_{\rm L L}(t)=\hat{p}_{\rm RR}(t)=e^{-u\, t}(1+u)(2+u(2+u))p_{\rm LL}(t)\, , \qquad \hat{p}_{\rm LR}(t)=\hat{p}_{\rm RL}(t)=e^{-u\, t}\frac{3 (u+1) (u (u+2)+2)}{2 (2 u (u+2)+3)}p_{\rm LR}(t)\, .
\end{equation}

By construction the matrix $\hat{R}=R$ and 
$$
R=\frac{1}{4}\begin{pmatrix}
1&3\\
3&1
\end{pmatrix}\, ;
$$ 
the positive eigenvector of this matrix is $v=(1,1)^T$ which is indicating that the activities $Q_{\rm L}=Q_{\rm R}$, as expected since the problem was originally symmetric and our modified probabilities also obeys that symmetry. To find the actual value of the activity one thus needs to choose $Q_{\rm L}=Q_{\rm R}=Q$ such that the empirical measure (as given by Eq.~\eqref{TiltEmp}) is normalized:
$$
\int d\psi {\mu}(\psi)= 2\, Q\, \int_0^\infty dt\, \hat{s}(t)=1\, ,
$$
where $\hat{s}(t)=\hat{s}_1(t)=\hat{s}_2(t)$ is the survival probability for the modified process, which is, again by symmetry, equal for both initial reset states. This provides the value of the single jump activity $Q$ as a function of $u$:
\begin{equation}
Q=\frac{\left(2 u^2+4 u+3\right) \left(u^3+3 u^2+4 u+2\right)}{2 \left(3 u^4+12 u^3+20 u^2+16 u+6\right)}\, .
\end{equation}
Substituing the $\hat{p}_{ij}(t)$ and the $\hat{c}_i$ into the functional \eqref{CostModProbab}, we also get the probability of this modified process as a function of $u$. We can thus plot, as reported in the figure in the main text, a plot of the "cost" of the process versus activity of the single jump in the modified process to obtain a bound to the rate function for a single jump activity. 

In order to obtain the same bound for the time-integrated entanglement we need to express the entantanglement entropy of the modified process as a function of $u$. 
The time-integrated entanglement entropy for the process can be obtained throught the empirical measure 
$$
\hat{\mathcal{S}}_\tau=\int d\psi\, {\mu}(\psi) S(\psi)=Q\sum_{i={\rm L,R}}\int_0^\infty dt\, \hat{s}_i(t)\, S(\varphi_{i t})\, ,
$$
where $S(\varphi_{i\, t})$ is the entanglement entropy for the state that has started from the reset state $i$ and has evolved with no jump occurring up to time $t$:
$$
S(\varphi_{{\rm L}t})=S(\varphi_{{\rm R} t})=-\cos ^2\left(\frac{t}{2}\right) \log \left(\cos ^2\left(\frac{t}{2}\right)\right)-\sin ^2\left(\frac{t}{2}\right) \log \left(\sin
   ^2\left(\frac{t}{2}\right)\right)\, .
$$
Thus again one has a parametric dependence of both the entanglement entropy and the probability of the modified process as a function of $u$ from which one can extrapolate the bound to the entanglement entropy rate function.

\end{document}